# Biodegradation of 2-ethylhexyl nitrate (2-EHN) by *Mycobacterium austroafricanum* IFP 2173

**Elodie Nicolau[1], Lucien Kerhoas[2], Martine Lettere[2], Yves Jouanneau[3], Rémy Marchal[1]**

[1] IFP, Département de Biotechnologie et Chimie de la Biomasse, 1-4 avenue du Bois Préau, 92852 Rueil-Malmaison Cedex, France

[2] INRA, Unité de Phytopharmacie et Médiateurs Chimiques, Route de St-Cyr, 78026 Versailles Cedex, France

[3] CEA, DSV, iRTSV, Laboratoire de Chimie et Biologie des Métaux, 17 rue des Martyrs F-38054 Grenoble ; CNRS, UMR5249, F-38054 Grenoble ; Université Joseph Fourrier F-38000 Grenoble, France.

Running title: Biodegradation of 2-ethylhexyl nitrate

Corresponding author : Rémy Marchal, IFP, Département de biotechnologie 1-4 avenue Bois-Préau, 92852 Rueil-Malmaison Cedex France ; E-mail: remy.marchal@ifp.fr; Phone : 33 (0)1 47 52 69 24.

**ABSTRACT**

2-Ethyhexyl nitrate (2-EHN) is a major additive of fuel which is used to comply with the cetane number of diesel. Because of its wide use and possible accidental release, 2-EHN is a potential pollutant of the environment. In this study, *Mycobacterium austroafricanum* IFP 2173 was selected among several strains as the best 2-EHN degrader. The 2-EHN biodegradation rate was increased in biphasic cultures where the hydrocarbon was dissolved in an inert non-aqueous phase liquid (NAPL), suggesting that the transfer of the hydrophobic substrate to the cells was a growth-limiting factor. Carbon balance calculation as well as organic carbon measurement indicated a release of metabolites in the culture medium. Further analysis by gas chromatography revealed that a single metabolite accumulated during growth. This metabolite had a molecular mass of 114 Da as determined by GC/MS and was provisionally identified as 4-ethyldihydrofuran-2(3H)-one by LC-MS/MS analysis. Identification was confirmed by analysis of the chemically synthesized lactone. Based on these results, a plausible catabolic pathway is proposed whereby 2-EHN is converted to 4-ethyldihydrofuran-2(3H)-one, which cannot be metabolised further by strain IFP 2173. This putative pathway provides an explanation for the low energetic efficiency of 2-EHN degradation and its poor biodegradability.

## INTRODUCTION

2-Ethyhexyl nitrate (2-EHN) is the nitric ester of 2-ethyl-1-hexanol. It is added at 0.05 % to 0.4 % to diesel formulations in order to boost the cetane number. As a result of the large use of diesel worldwide, the 2-EHN market is about 100, 000 tons per year.

Although biodegradability has for a long time been regarded as a relevant characteristic of chemicals, it was only recently incorporated to safety assessments. Considering fuel oils, large volumes of oxygenates such as MTBE have been added to gasoline since 1992 (19). Because of lack of knowledge on their biodegradability and insufficient safety regulation, pollution cases resulting from accidental releases occurred in many countries. In the US for example, as many as 250 000 sites may have been polluted from leaking underground fuel tanks (36). Poor knowledge of the biodegradation of widely-used chemicals may also hide specious concerns relating to the toxicity of metabolic products. For example, degradation of chlorinated aromatics such as 4-chlorocatechol in soil gave rise to the formation of an antibiotic, protoanemonin, which is detrimental to soil microcosms (6).

In case of accidental release of 2-EHN into the environment, the fate and impact of the pollution are unpredictable because of the scarcity of data on 2-EHN biodegradation. Screening tests have been recommended by both the U.S. Environmental Protection Agency (35) and the OECD (24) to evaluate the biodegradability of commercial substances. In this context, the so-called criterion of "ready biodegradability" requires that the tested substance be biodegraded to a level of 60 % within 28 days (5). Standard degradation experiments showed that 2-EHN could not be considered readily biodegradable (34). It was assumed that 2-EHN was poorly available to microbial communities because of its low water solubility and its high volatility.

In fact, 2–EHN displays both a low vapour pressure corresponding to about 1.9 mg/l at 20°C and a moderate solubility in water (12.6 mg/l at 20°C). Therefore, 2-EHN is expected to form a separate organic phase in aqueous solution even when present in low amount. 2-EHN is also a rather hydrophobic molecule as indicated by a log $K_{o/w}$ value of 5.24. Hydrophobic compounds with log $K_{o/w}$ values in the range 1-5 are often toxic to cells because they insert into the lipid bilayer of the cell membrane, disturbing its integrity and causing cell permeabilization (13, 22).

The backbone of 2-EHN is a branched alkane, a type of molecules that is more resistant to biodegradation than linear alkanes. The metabolism of both linear and branched hydrocarbons by bacteria involves enzymes of the β-oxidation pathway (3). In the case of branched alkanes, their degradation may lead to the formation of β-substituted acyl-CoA intermediates that block β-oxidation (27). Such a metabolic blockage has been encountered during the degradation of terpenoids such as citronellol, geraniol and nerol (10, 28). If a quaternary carbon atom occurs at the end of an alkane chain, the result is a molecule quite resistant to microbial attack (18).

In a recent study, microbial communities endowed with the ability to degrade 2-EHN were obtained by enrichment from activated sludge or soil samples (33). The isolation of pure strains able to utilize 2-EHN as sole source of carbon and energy proved rather difficult. Nevertheless, among several strains of fast-growing Mycobacteria previously isolated on other hydrocarbons, some strains, all identified as *Mycobacterium austroafricanum,* were found to degrade 2-EHN.

In the present study, the kinetics of 2-EHN degradation by selected strains was investigated. *M. austroafricanum* IFP 2173, which showed the highest rate of degradation, was chosen for further investigation of 2-EHN catabolism. As a means to reduce the expected toxic effect of 2-EHN on bacterial cells and increase its bioavailability in aqueous media, bacterial cultures were mostly carried out in biphasic media. Such biphasic cultures, including a non aqueous phase liquid (NAPL) that serves as solvent for the hydrophobic substrate have already been implemented to facilitate the degradation of various toxic or recalcitrant compounds (2, 4, 7, 12, 25, 26). A metabolite that accumulated during growth was detected in the culture medium and identified by LC-MS/MS. Based on our data, a plausible pathway for 2-EHN catabolism by *M. austroafricanum* IFP 2173 is proposed.

## MATERIALS AND METHODS

### Microorganisms and culture conditions

The strains used in this study were *M. austroafricanum* IFP 2173 (30), isolated on iso-octane, *M. austroafricanum* IFP 2012 (11) and *M. austroafricanum* IFP 2015 (15) both isolated on MTBE, *M. austroafricanum* C6 (14), *M. austroafricanum* Spyr_Ge_1 and *M. austroafricanum* BHF 004 (J. C. Willison, unpublished data), all isolated on pyrene.

The culture medium consisted of a mineral salts solution (8) supplemented with 0.1 g/l of yeast extract. The carbon source was added after medium sterilization (120°C for 20 min). All cultures were incubated at 30°C with shaking (150 rpm).

### Chemicals

2-EHN (CAS Number 27247-96-7), 2-ethyhexanol, 2-ethylhaxanoic acid, MTBE, decahydronaphtalene, 3-methyldihydrofuran-2(3H)-one, $Et_2Zn$, and HMN were obtained from Sigma Aldrich (Saint Quentin Fallavier, France). Mineral salts were from VWR (Fontenay-sous-Bois, France).

### Biodegradation experiments

Biodegradation tests were performed in 120-ml flasks closed with Teflon-coated stoppers and sealed with aluminium caps. Unless otherwise indicated, 4.8 mg of 2-EHN (or 2-ethylhexanol or 2-ethylhexanoic acid) was added to 10 ml of the medium supplemented with 500 µl of 2,2,4,4,6,8,8-heptamethylnonane (HMN). Cultures were adjusted to an optical density ($O.D._{600}$) of 0.2 using washed pellets of centrifuged precultures grown on Tween 80 (2.5 g/l) as sole source of carbon. The degradation rate was monitored by measuring at regular intervals the $CO_2$ evolved in the headspace by gas chromatography (GC). Residual 2-EHN was estimated as described below in triplicate. Abiotic controls were supplemented with mercuric chloride (0.2 mg/l) and endogenous controls, lacking a carbon source but containing HMN, were performed under similar conditions.

**Analyses of substrate and products**

Culture grown on 2-EHN were filtered on a PTFE membrane (0.45 µm) and cell biomass was determined as dry weight after lyophilisation of the cell pellet. When HMN was omitted from the growth medium, the total organic carbon (TOC) was measured on the filtrates using a TOC-5050 carbon analyser (Shimadzu) according to the European norm NF EN 1484. Residual 2-EHN in the culture filtrate, as well as derived metabolites, were extracted with 10 ml of methyl-*tert*-butyl ether (MTBE) containing 0.05 % (v/v) of decahydronaphthalene as internal standard. After 30 min of shaking and static overnight incubation at 4°C, the solvent extracts were analysed by GC with flame ionization detection (FID). A Varian 3400 chromatograph (Sugarland, USA) equipped with a CP-Sil Pona CB column (0.25 mm by 50 m) obtained from Chrompack (Raritan, NJ) was used. The carrier gas was helium. The temperature of the injector and the detector were set at 250 and 280°C, respectively. The column temperature was varied from 100°C to 200°C at 4°C/min, then from 200°C to 259°C at 20°C/min**.**

Time courses of 2-EHN degradation and metabolite excretion were performed in flasks which were sacrificed at regular time intervals. $CO_2$ in flask head space was measured with a Varian 3400 gas chromatograph (Sugarland, USA) equipped with a catharometric detector and a PorapackQ (80/100 mesh, 2m) (Chrompack, Raitan, NJ). The net amount of $CO_2$ produced was determined as the difference between the final quantity found in the test flasks and that found in hydrocarbon-free flasks.

**Kinetics of $O_2$ consumption**

Continuous monitoring of substrate oxidation was carried out through measurement of $O_2$ consumption using a respirometer (Sapromat D12-S, Voith, Germany). Flasks containing 250 ml of culture medium and 125 µl of 2-EHN as carbon source were inoculated with *M. austroafricanum* IFP 2173 to an optical density (O.D.$_{600}$) of 0.1. Incubation was carried out at 30°C with shaking in the presence or absence of HMN (12.5 ml). Cultures and substrate-free controls were performed in triplicate.

**Chemical synthesis of 4-ethyltetrahydrofuran-2(3H)-one**

4-EDF was synthesized according to a published procedure (1). In a three-necked flask containing dry toluene (5 ml), $Cu(OTf)_2$ (0.025 mmol) and $P(OEt)_3$ (0.05 mmol) were successively added. The mixture was stirred for 30 min at room temperature to obtain a colourless solution. After cooling to -20°C, Zinc di-ethyl (5 mmol previously dissolved in hexane) was added followed by furan-2(5H)-one (5 mmol). The reaction was allowed to warm to 0°C for 6 h, then incubated at room temperature and monitored by GC. After completion of the reaction, the mixture was hydrolysed with aqueous 5N HCl, then extracted with diethyl ether (2 x 15 ml); the organic phase was dried over $MgSO_4$ and concentrated *in vacuo*. The crude product was purified by column chromatography on $SiO_2$ using a mixture of diethyl ether / pentane: 80/20) as eluent.

**Coupled MS analyses**

GC-MS analysis was carried out under chromatographic conditions identical to those described above for GC-FID. Mass spectra were acquired in the split mode with a time of flight mass spectrometer (Tempus TOF MS, Thermo Finnigan).

LC–MS–MS was performed using an HPLC system (Alliance 2695, Waters, Guyancourt, France) coupled to a Quattro LC triple quadrupole mass spectrometer (Micromass,

Manchester, UK) with an electrospray interface. Data were acquired in the positive or negative ionization modes and processed with MassLynx NT 4.0 system. The electrospray source voltages were: capillary 3.2 kV, extractor 2 V, cone voltage 22 and 17 V under positive mode, respectively. The source block and desolvation gas were heated at 120°C and 350°C, respectively. Nitrogen was used as nebulisation and desolvation gas (75 and 350 $l\ h^{-1}$ respectively). For MS–MS, collisional induced dissociation (CID) was performed under argon ($2.5\ 10^{-3}$ mbar) at a collision energy set between 10 and 40 eV.

## RESULTS

### Time course of 2-EHN biodegradation by selected strains

Kinetics of 2-EHN biodegradation was studied using a few bacterial strains previously selected among environmental isolates and collection strains for their ability to attack this compound (33). Most of these strains were identified as members of the *Mycobacterium* genus. In order to avoid growth inhibition due to 2-EHN toxicity, HMN was added as NAPL to the bacterial cultures, and biodegradation time courses were monitored by measuring the $CO_2$ production in the culture headspace. Biodegradation kinetics were found to vary widely depending on bacterial strains (data not shown). *M. austroafricanum* IFP 2173 was the fastest and most efficient of the microorganisms tested since it produced the largest amount of $CO_2$ (37 µmol per flask) after 13 days of incubation. *M. austroafricanum* IFP 2173 was also the only strain able to grow on 2-EHN in the absence of HMN (data not shown).

**Effect of 2-EHN supply mode on the biodegradation rate**

The impact of NALP addition on 2-EHN biodegradation by strain IFP 2173 was studied through continuous monitoring of substrate-dependent oxygen consumption by respirometry. In the culture lacking HMN, $O_2$ uptake started after a lag phase of about one day, then increased with time according to a sigmoidal curve (Fig. 1). The maximal growth rate ($\mu_{max}$) could be deduced from oxygen uptake rate assuming that the biomass yield remained constant during growth. Over a 9-day period of growth, $\mu_{max}$ was calculated to be 0.29 $day^{-1}$ on average, corresponding to a generation time of 2.4 days. In the HMN-containing culture, the lag phase was shorter and the $O_2$ uptake became linear after a very short exponential phase ($\mu_{max}$ =0.29 $day^{-1}$). The maximal rate of $O_2$ uptake was 5.3 mmol/day, and the overall $O_2$ consumption reached a maximum of 2.9 mmol, compared to 2.6 mmol for cells grown without HMN.

The effect of 2-EHN concentration on growth was studied in HMN-containing cultures (Fig. 2). The concentration of 2-EHN had little effect on the specific growth rate. During the linear phase of growth, the $O_2$ uptake rate increased proportionally to the 2-EHN concentration in the culture medium up to 3 g/l. This indicated that the 2-EHN diffusion rate from HMN to the water phase was a limiting factor for bacterial growth. At 2-EHN concentrations higher than 3 g/l, bacterial growth was inhibited as indicated by both slower oxygen uptake rates and lower overall $O_2$ consumption. For 2-EHN concentrations lower than 3 g/l, no residual substrate was detected in the culture medium by the end of growth and the $O_2$ consumption was roughly proportional to the amount of substrate supplied.

**Carbon balance of 2-EHN biodegradation by *M. austroafricanum* IFP 2173**

In order to determine the carbon balance of 2-EHN biodegradation, *M. austroafricanum* IFP 2173 was cultivated in mineral medium lacking HMN to avoid perturbation of TOC measurements by HMN. The culture was stopped when no more $CO_2$ was released, which coincided with the total consumption of 2-EHN (see Fig. 4). The biomass formed, the TOC in the filtered culture medium and the amount of $CO_2$ released were measured. The carbon recovery as metabolites and cell biomass was calculated by taking into account the elementary compositions of substrate and products (Table 1). A carbon recovery rate of 92 % was obtained for the 2-EHN bioconversion. Carbon converted into biomass (94 mg/l) and $CO_2$ (165 mg/l) amounted together to only 33 % of the total carbon produced. Accordingly, a high proportion of the substrate-derived carbon was recovered in the clarified culture medium (67 %), possibly reflecting metabolite accumulation.

**Identification of a metabolite excreted in the culture**

GC-FID analysis of culture fluid extracts performed during 2-EHN degradation experiments revealed the gradual increase in concentration of an unknown compound with a retention time shorter than that of 2-EHN. This finding suggested that a metabolite might have accumulated during growth and accounted for the substantial level of TOC previously detected in the supernatant of 2-EHN grown cultures. High resolution mass spectral analysis of this compound (Fig. 3 a) showed that it had a molecular mass of 114.07 Da and the following chemical formula: $C_6H_{10}O_2$. The mass spectrum of this compound did not match any of the spectra currently available in the databases. Nevertheless, a comparison of the LC-MS-MS data of the excreted product with those of 3-methyldihydrofuran-2(3H)-one, a commercially-available product, revealed several common fragment ions. The analysis also indicated that the molecule did not contain any carboxylic or hydroxyl groups (Fig. 3 b). Taken together, our data indicated that the product of interest might be 4-ethyldihydrofuran-2(3H)-one (4-EDF), which can also be designated as β-ethyl-γ-butyrolactone. In order to confirm the structure of the metabolite, the chemical synthesis of 4-EDF was undertaken as described under Materials and Methods (1). The LC-MS-MS characteristics of the synthesized lactone were identical to those of the metabolite, confirming that the product which accumulated in cultures of *M. autroafricanum* IFP 2173 grown on 2-EHN was 4-EDF.

The rate of 4-EDF accumulation was assessed by GC-FID analysis of the culture fluid during growth. Fig. 4 shows that 4-EDF formation and $CO_2$ release were concurrent with 2-EHN degradation.

**Biodegradation of 2-EHN-derived compounds**

As a means to elucidate the biodegradation pathway of 2-EHN by *M. austroafricanum* IFP 2173, we tested compounds with structures derived from 2-EHN as possible substrates. 2-ethylhexanol, the primary alcohol resulting from 2-EHN hydrolysis, was biodegraded, yielding 2-ethylhexanoic acid and 4-EDF. 2-ethylhexanoic acid, the product resulting from 2-ethylhexanol oxidation was not biodegraded, even in the presence of HMN. This compound is considered to be toxic for most bacteria (21). It should be noted that 2-EHN can be used as sole nitrogen source by strain IFP 2173, indicating that nitrate is formed, probably as a result of an initial attack on 2-EHN by an esterase (data not shown). 2-EHN biodegradation was also tested in the presence of isooctane, the compound on which *M. austroafricanum* IFP 2173 was selected. Diauxic growth was observed, the strain degrading isooctane first and then 2-EHN into 4-EDF (data not shown).

**DISCUSSION**

2-EHN is a recalcitrant compound which was considered not readily biodegradable according to standard procedures (34). However, we demonstrated in the present study, that selected strains of Mycobacteria were able to slowly utilize 2-EHN as sole source of carbon under defined culture conditions. The poor biodegradability of 2-EHN might be the consequence of two factors, first the low occurrence of micro-organisms able to use it as carbon source, and second its inhibitory effect on bacterial growth even at low concentration. 2-EHN inhibition was illustrated by the experiment described in Fig. 2, and by the lack of growth of all strains tested in HMN-free cultures, except *M. austroafricanum* IFP 2173. This strain, isolated for its ability to degrade isooctane, a branched alkane (31), demonstrated wide capabilities for hydrocarbon biodegradation (16, 32). Like many members of the *Corynebacterium-Mycobacterium-Nocardia* (CMN) group of Gram-positive bacteria, it may be resistant to toxic hydrocarbons thanks to the properties of its cell envelope, which is highly rigid and contains mycolic acids (29). In *Mycobacteria*, mycolic acids are very long fatty acids ($C_{60}$-$C_{90}$) that contribute up to 60 % to the cell wall (9). The specific cell wall composition of the *M. austroafricanum* strains studied here probably accounts for their resistance to 2-EHN.

However, it is unclear whether the unique ability of strain IFP 2173 to grow on 2-EHN without NAPL is due to a cell wall composition slightly different from that of other strains or to some other strain-specific trait.

Biphasic cultures, involving addition of an inert NAPL like HMN was found to be critical for 2-EHN biodegradation and bacterial growth. In the HMN-free cultures, the dissolved fraction of 2-EHN represented only a minor part of the substrate supplied since it partitioned into three distinct phases *i.e.* the gas phase, the aqueous phase, and the bulk of insoluble 2-EHN. During the biodegradation process, the uptake of dissolved substrate was counterbalanced by the equilibrium transfer of 2-EHN from the bulk of substrate ($S_{subNAPL}$) to the aqueous ($S_{aq}$) according to the following scheme:

$$S_{sub/NAPL} \underset{}{\overset{T_{NAPL/aq}}{\rightleftharpoons}} S_{sub/aq} \xrightarrow{\text{Growth phase}} X$$

where $S_{subNAPL}$ and $S_{aq}$ represent the amounts of substrate in the bulk and in the aqueous phase, respectively, X is the cell biomass and $T_{NAPL/aq}$ is the substrate transfer rate of 2-EHN to the culture medium. In HMN-containing cultures, the dissolved 2-EHN was mainly confined to NAPL. Because of the high hydrophobicity of their cell walls, microbial cells tightly adhered to NAPL and direct contact was thus the most probable mode of substrate uptake (7, 12). Accordingly, the large NAPL volume (500 µl of HMN versus 5 µl of 2-EHN in the case of the NAPL-free culture), which increased substrate bioavailability, probably accounted for its higher efficiency of assimilation by the microorganisms. Such conditions of substrate delivery were apparently required to promote growth on 2-EHN of *M. austroafricanum* strains other than strain 2173.

The biodegradation of 2-EHN by *M. austroafricanum* IFP 2173 illustrates the remarkable metabolic capabilities of this stain towards recalcitrant hydrocarbons. Indeed, it can degrade another methyl branched alkane, 2,2,4-trimethylpentane (31), suggesting that it produces enzymes specific for the degradation of anteiso-alkanes. Nevertheless, our results indicate that degradation of 2-EHN by strain IFP 2173 is partial, and gives rise to the release of an acyl with an ethyl substituent in the beta position. At least two reasons might explain the accumulation of this metabolite: i) strain IFP 2173 lacks enzymes able to degrade it, ii) because of the ethyl group in beta position, the metabolite might block the enzyme catalysing the next step in the degradation of branched alkanes.

Considering the high biodegradation potential of strain IFP 2173, we recently observed that this strain can degrade other xenobiotic compounds structurally related to 2-EHN such as bis(2-ethylhexyl)phthalate (data not shown) used as plasticizer (21, 23). The biodegradation of this compound by *Mycobacterium* sp. NK0301 has been reported (20). This bacterium utilized phthalate as carbon and energy source and left the carbon skeleton of the 2-ethylhexyl moiety intact, releasing it as 2-ethylhexanol or 2-ethylhexanoic acid. In comparison, strain IFP 2173 degraded bis(2-ethylhexyl) phthalate and utilized the 2-ethylhexyl moiety, achieving a higher degree of degradation (data not shown).

The biodegradation of 2-EHN by strain IFP 2173 gave rise to the accumulation of a lactone which was identified as 4-EDF. The lactone formed by cyclization of a breakdown product, a branched pentanoic acid, which was not metabolized further by the bacteria. The partial degradation of 2-EHN certainly explains the observed slow growth ($\mu_{max}$ = 0.29 day$^{-1}$) and poor growth yield of cultures utilizing this compound as sole C source.

Considering the structure of the intermediate metabolite and the known degradation pathway of *n*-alkanes (18), we propose for the first time a plausible metabolic pathway for 2-EHN degradation (Fig. 5). The pathway would start by a simultaneous or sequential attack of the molecule on both extremities, with an esterase activity hydrolyzing the nitric ester bond and an oxygenase catalyzing the hydroxylation of the distal methyl group. The involvement of an esterase that would release nitrate was inferred from the observation that strain IFP 2173 utilized 2-EHN as nitrogen source. The existence in this strain of an hydroxylase active on branched alkane is expected since it grows on isooctane (31). The intermediate metabolite that would form, 2-ethylpentan-1,5-diol, is proposed to be oxidized to a carboxylic acid in two steps involving successively an alcohol and an aldehyde dehydrogenase. After activation by coenzyme A, the resulting 5-(hydroxymethyl)heptanoic acid would undergo one cycle of classical β-oxidation to give 3-(hydroxymethyl)pentanoic acid, which would spontaneously convert to 4-EDF by cyclisation. Since the substrate underwent a single turn of β-oxidation only two carbon atoms (out of eight in 2-EHN) could reach the TCA cycle, accounting for the low percentage of carbon released as $CO_2$ (12%).

The proposed pathway now needs to be assessed experimentally by identifying enzymes involved in 2-EHN degradation. To this end, we have undertaken a proteomic analysis to find out the proteins that are induced upon incubation of strain IFP2173 with 2-EHN.

**ACKNOWLEDGEMENTS**

This work was supported by a Convention Industrielle de Formation par la Recherche (CIFRE) fellowship from the Association Nationale de la Recherche Technique (ANRT) to E. Nicolau and grants from the IFP. We thank F. Léglise for helpful discussions and J. C. Willison for critical reading of the manuscript.

**Table 1**: Carbon balance of 2-EHN biodegradation by *M. austroafricanum* IFP2173

Cultures (10 ml) were performed at 30°C in 120-ml flasks.

| Substrate or product | Mass change[a] (mg/l) | Carbon balance | |
|---|---|---|---|
| | | Carbon change[a] (mg/l) | Carbon recovery (%) |
| 2-EHN | 482 | 269 | 0 |
| Cell biomass | 94 | 50[b] | 19 |
| $CO_2$[c] | 115 | 31 | 12 |
| TOC[d] | 165 | 165 | 61 |
| Total products | | | 92 |

[a] Considering the whole content of the culture flasks.

[b] Carbon to dry biomass ratio was assumed to be 52 % (17). Dry biomass was determined from 100 ml cultures grown in 1-L flasks.

[c] $CO_2$ was determined after acidification of the culture

[d] Total organic carbon (TOC) measured in the culture fluid after filtration through a 0.22 µm membrane.

## Figure legends

**Fig 1.** Effect of a non aqueous liquid phase (HMN) on the rate of oxygen consumption by *M. austroafricanum* IFP 2173.

Cultures (250 ml) were grown in the flasks of a respirometer and contained 125 µl of 2-EHN as carbon source. Cultures were incubated in the presence (black line) or absence (grey line) of HMN (12.5 ml).

**Fig. 2.** Effect of 2-EHN concentration on oxygen consumption by *M. austroafricanum* IFP 2173**.**

Biphasic cultures contained a variable concentration of 2-EHN and 12.5 ml of HMN. Maximal rates of $O_2$ uptake or Vmax (■) and overall $O_2$ consumption (♦) were determined.

**Fig. 3.** MS characterization of the metabolite produced by strain IFP 2173 upon degradation of 2-EHN.

**a.** High resolution electron impact mass spectrum of the accumulated metabolite as obtained by GC-MS analysis.

**b.** CID/MS/MS product ion spectrum of the protonated molecule (MH+) obtained by LC-MS/MS analysis at a collision energy of 10 eV.

**Fig 4.** Accumulation of 4-EDF during 2-EHN biodegradation.

Parallel cultures were carried out in 120-ml flasks and removed at the times indicated for extraction and measurements of 2-EHN (♦) and 4-EDF (▲). $CO_2$ (■) was determined in a separate culture flask. Residual 2-EHN is the fraction of hydrocarbon which stayed bound to the flask wall and stopper, and remained inaccessible to bacteria.

**Fig 5.** Proposed pathway for 2-EHN biodegradation by *M. austroafricanum* IFP 2173.

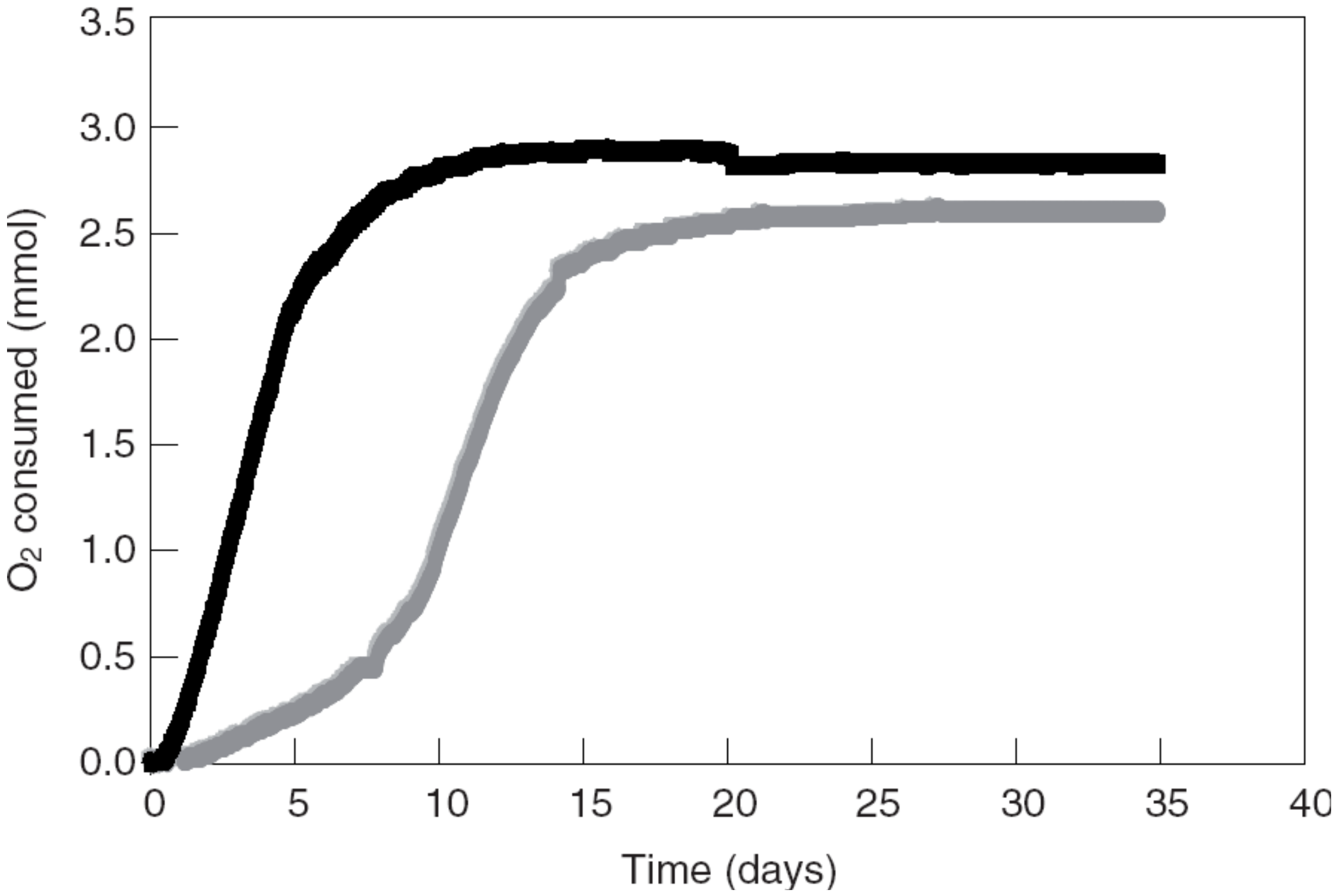

Fig. 1

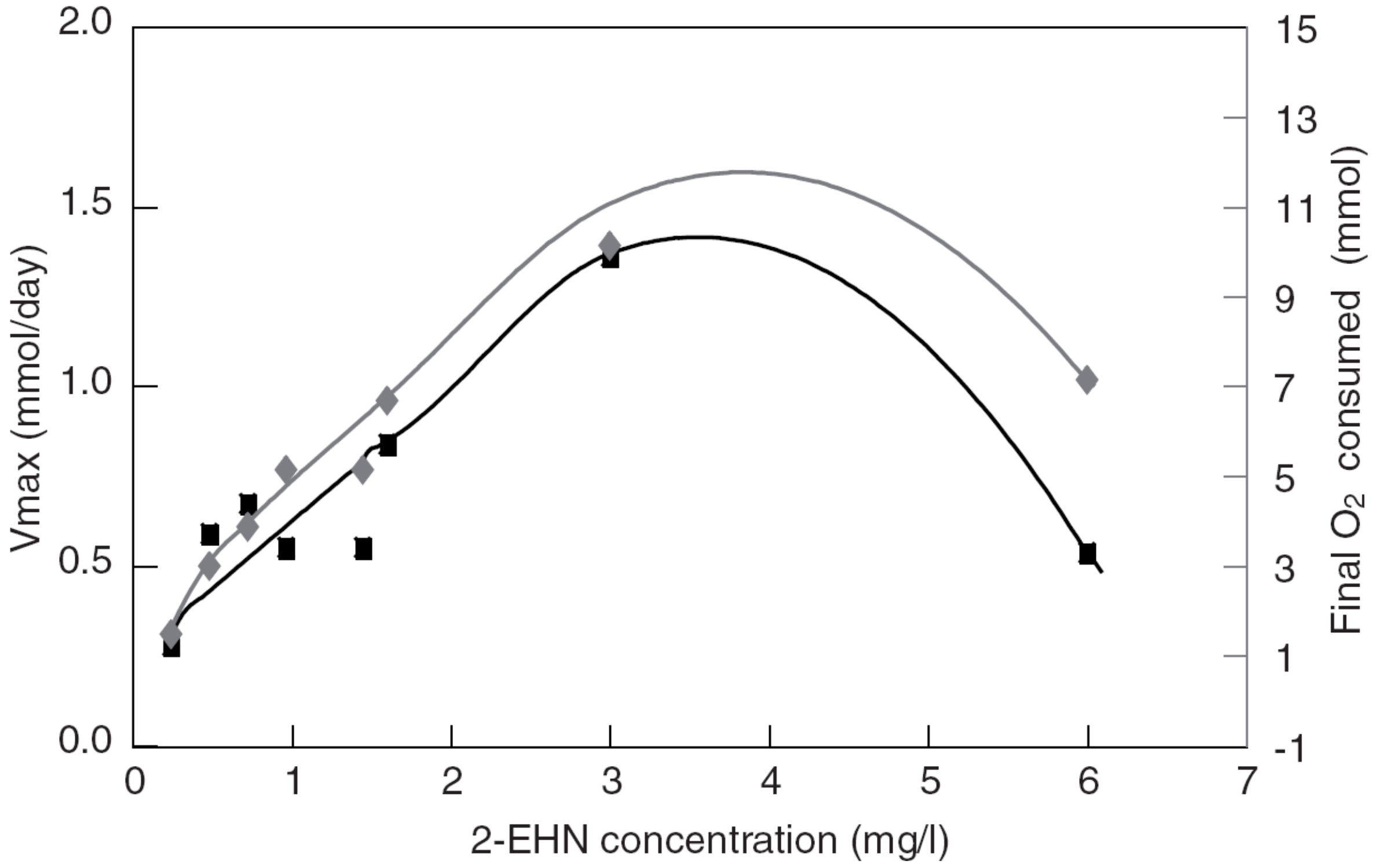

Fig. 2

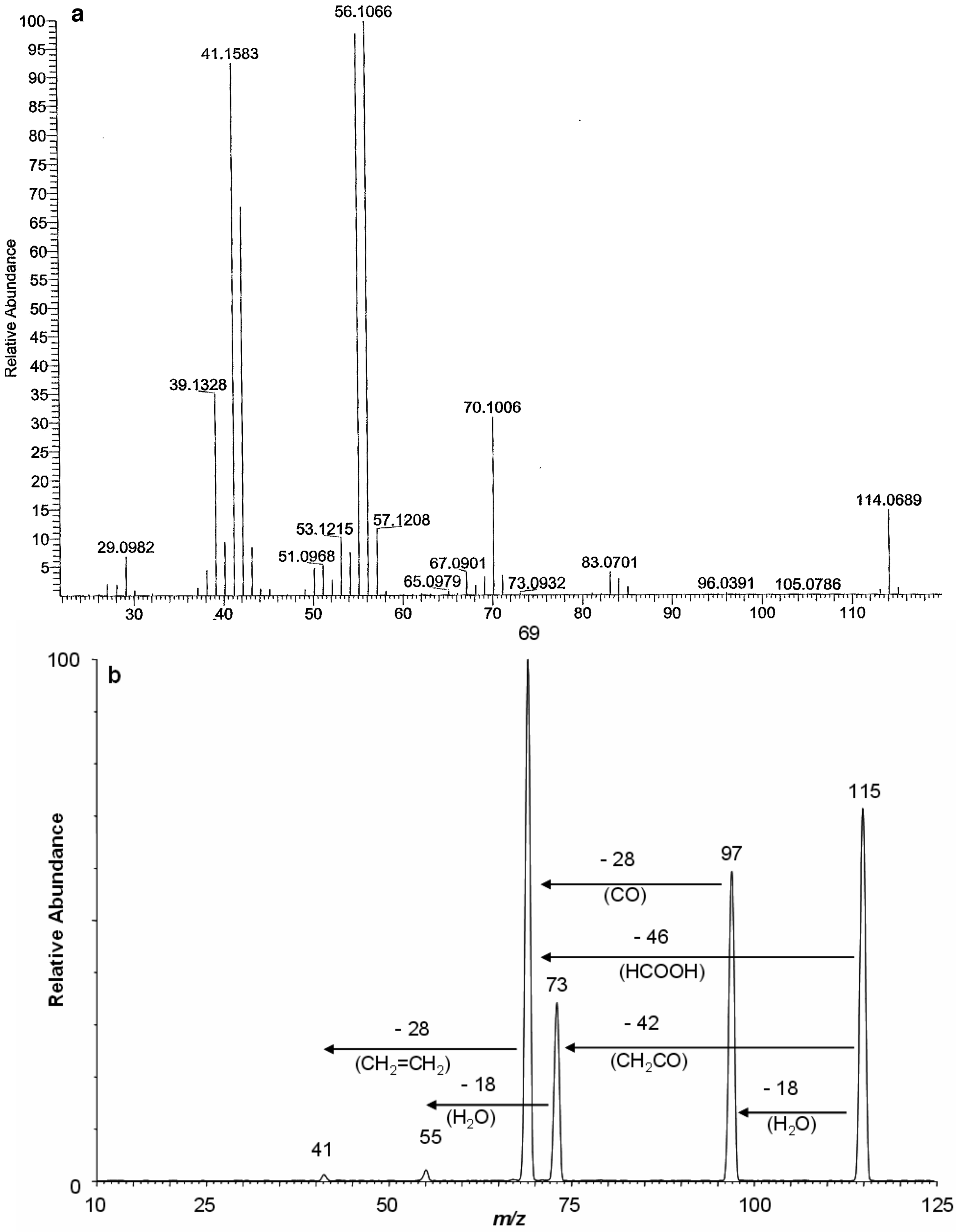

Fig. 3

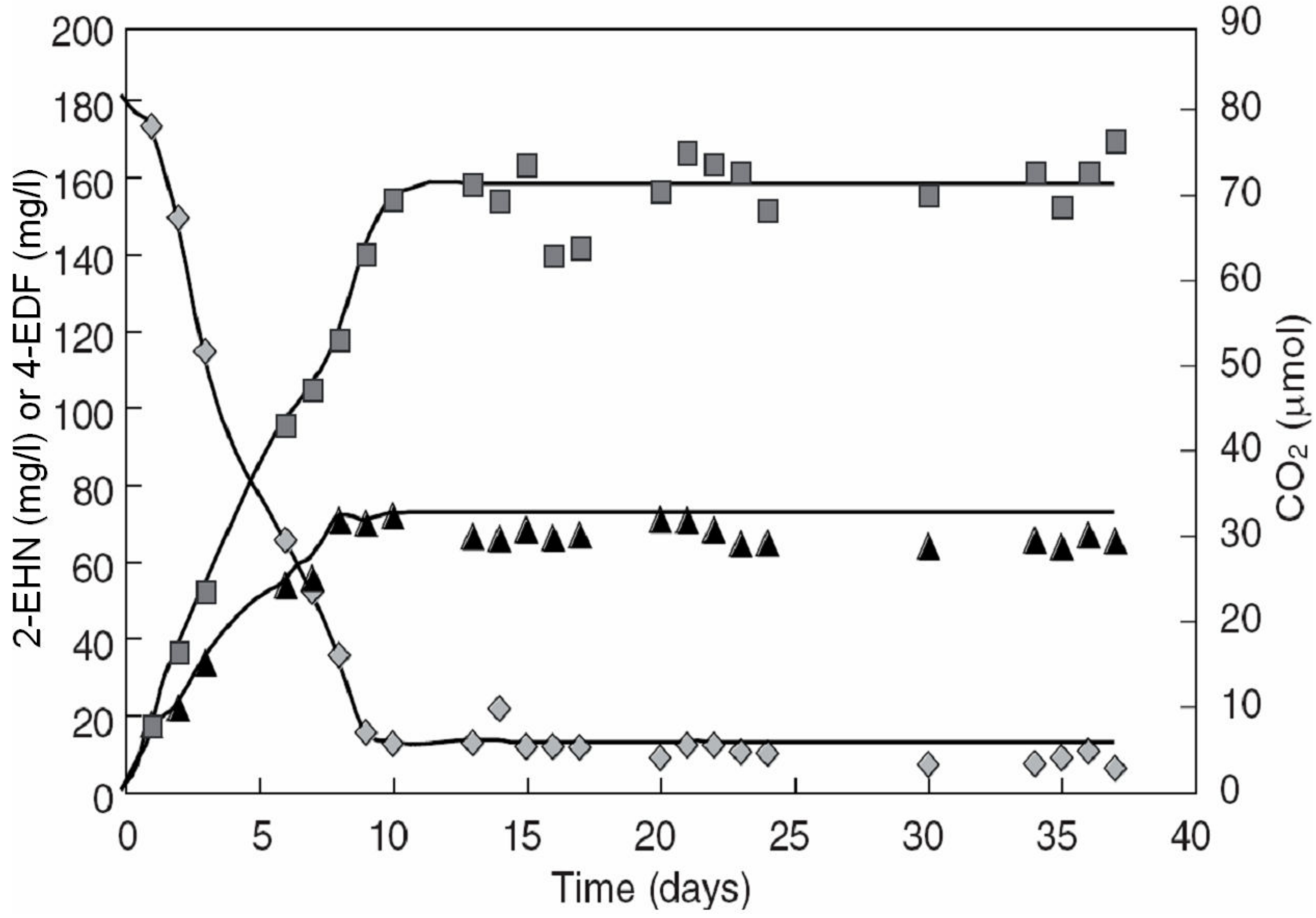

Fig. 4

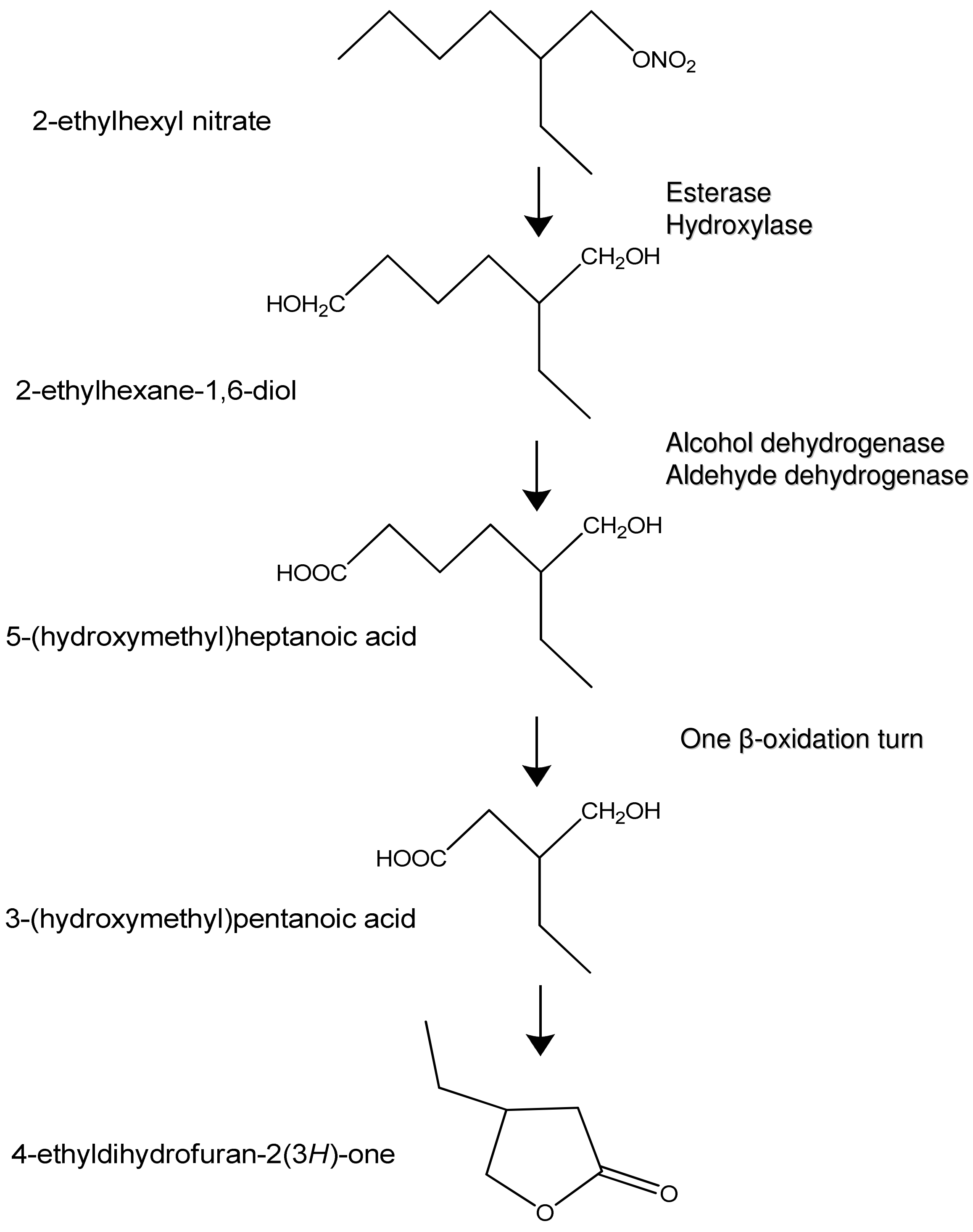

Fig. 5